\documentclass[prl,aps,preprint]{revtex4}

\usepackage{graphicx}
\usepackage{amsmath}
\usepackage{amstext}

\begin{document}

\title{Light driven structuring of glasses}
\date{\today}

\author{B. P. Antonyuk\thanks{Electronic mail: antonyuk@isan.troitsk.ru},  A.Z. Obidin* , and K.E. Lapshin* }

\begin{abstract}
Theoretical and experimental evidence of light driven structuring of glasses is
presented. We show that light overcomes Coulomb repulsion and effective
electron-electron interaction in glasses under strong light pumping becomes
attractive. As the result homogenious distribution of trapped electrons gets
unstable and macroscopic electron bunches are grown. At different conditions
ordered structures with period $2 \mu m \pm 0.2 \mu m$ determined by internal
properties of the material are formed These structures were observed in
ablation: surface profile after laser treatment reveals ordered pattern
corresponding to the light induced electron domains.

\end{abstract}

\affiliation{Institute of Spectroscopy, Russian Academy of Science, 142190
Troitsk, Moscow Region, Russia. E-mail: Antonyuk@isan.troitsk.ru *Physics
Instrumentation Center, General Physics Institute, Russian Academy of Science,
142190 Troitsk, Moscow Region, Russia.}

\maketitle

\section{Introduction}

Direct and obvious method of light driven structuring is the using of a special
light field configuration. Generation of the light intensity grating and
corresponding grating of excitations are widely discussed. More interesting and
physically rich is structuring due to light driven self-organization. This
ordering belongs to the class of self-organization phenomena in open
dissipative systems and it is presented here. The typical examples of this
class are Benard convection \cite{Benard}, Belousov-Zhabotinsky reactions
\cite{Belousov} and Turing instability \cite{Turing}. All self-organization
phenomena are driven by some external flow through the system: heat in Benard
convection, chemical reagents in Belousov-Zhabotinsky reactions, etc. Optical
analogue of Turing instability was presented in \cite{Arecchi}. In the latter
case light amplitude has been modulated in time and space, which was crucial
for the organization to occur. Our investigations(see \cite{AntUFN},
\cite{OptComm}, \cite{Monograph} and references therein) have shown that steady
light flow through a system can provide ordering either. This light is a
driving force which bunches the electrons in the case in hand. An electron in
disordered media (we shall call them "glasses") moves in random potential
forming its energy spectrum. It consists of the extended electron and hole
bands separated by the "gap" filled by the local states. We shall discuss here
the case when photon energy is less than the gap in glass and light generates
transitions between local states only. Transitions between different traps are
of a spacial interest. They are responsible for the light induced current and
the structuring of the matter discussed. The electron transfer from an initial
trap $i$ to the final trap $f$ may be treated as generation of electron and
hole at sites $f$ and $i$ respectively. Strong Coulomb electron-hole
interaction is changed considerably with the shift of the distance between
traps $f$ and $i$ therefore strong electron-phonon coupling takes place. Light
absorption in the case of strong electron-phonon interaction is well studied.
The rate $w_{fi}$ of the electron transition between sites $i$ and $f$ is given
by the formula \cite{Kub}
\begin{equation}\label{w1}
w_{fi} = I \sigma_{0} cos^{2} \theta_{fi} exp(-\frac{(\hbar \omega -\epsilon_{fi} -A)^{2} }{\Delta^{2} } -\kappa_{fi} R_{fi}) \quad (\epsilon_{fi}>0),
\end{equation}
where $I$ is photon flux, $\hbar \omega$ is photon energy, $\epsilon_{fi} =
\epsilon_{f} - \epsilon_{i}$ is the energy of the electron excitation
(difference between electron energies $\epsilon_{f}$ in trap $f$ and
$\epsilon_{i}$ in trap $i$ ); $\bf{R_{fi}}={\bf{R}}_{f}-{\bf{R}}_i$;
$R_{fi}=|\bf{R_{fi}}|$, where vectors ${\bf{R}}_{f}$ and ${\bf{R}}_{i}$ points
positions of the traps; $\theta_{fi}$ is the angle between ${\bf R}_{fi}$ and
the light polarization vector, $\sigma_{0}\approx 10^{-18}cm^2$, $\Delta\approx
10^{-1}\div 10^{-2}eV$, $A\approx 10^{-1} \div 10^{-2}eV$ is the Stokes shift.
The light driven electron transitions in hand are accompanied by phonon
generation. Phonons are excluded from the formula (\ref{w1}) therefore exact
energy conservation law $\hbar \omega = \epsilon_{fi} + \hbar \Omega$ ($\hbar
\Omega$ is phonon energy) looks like conservation with the accuracy $\Delta
\approx \hbar \Omega$: according to (\ref{w1}) an electron may be transfered to
different levels in the range of the band width $\Delta$. Absorption band
(\ref{w1}) as a function of $\epsilon_{fi}$ gains maximum (resonance) at the
energy $M = \hbar \omega -A$. We shall call the excited states with the
$\epsilon_{fi}<M$ as low energy excitations (deep levels) and excited states
with the $\epsilon_{fi}>M$ as high energy ones.

\section{Light driven bunching of electrons}

As a first step we shall examine light driven kinetics of a single electron in
external electric field $\bf{E}$ . It is nontrivial: as we shall see electron
is shifted by light predominantly in opposite to the electric force $\bf{F} =
e\bf{E}$ direction. The idea of such behavior is following. If vector of
electron transfer is ${\bf{R}} = {\bf{R}}_f - {\bf{R}}_i$ then dipole moment
generated is ${\bf{d}} = e\bf{R}$ ($e < 0$ is electron charge). Field
contribution to the energy of the excitation is $\delta \epsilon_{fi} = -
({\bf{d}\bf{E}}) = - e({\bf{R}\bf{E}}) \equiv -(\bf{R}\bf{F})$ . If the
electron is transported in opposite to the electric force direction
($(\bf{R}\bf{F}) < 0$)  the energy of the electron excitation $\delta
\epsilon_{fi}$ becomes larger; low energy excitation becomes closer to
resonance $M$ and therefore probability of its generation increases. This means
that electron transfer to the deep levels goes predominantly in opposite to the
electric force direction. Electron transitions to high levels go mainly in the
direction of the force but these levels have less life time in comparison with
deep levels therefore contribution of the deep levels prevails. In the case of
two electrons according to the above idea light would push each electron in the
direction of the other electron. Let the second electron is embedded at
positions $j$ then Coulomb interaction shifts all energy levels for the first
electron  and, hence, shifting the excitation energies by
$e^2/R_{fj}-e^2/R_{ij}$ for the transition $i \longrightarrow f$ of the first
electron. Transitions which reduces electron-electron distance give positive
contribution to the Coulomb energy and shift deep excitations towards the
resonance $M$. That is why low energy transitions of an electron are directed
preferably towards the other electron and effective electron-electron
interaction becomes attractive under strong pumping. Due to shorter life time
high energy excitations play minor role in comparison with deep levels and
therefore contribution of low energy excitations dominates. At high pumping
these transitions dominate also over ordinary mobility which is quite low and
independent on laser power.  Effective attraction between electrons makes their
homogeneous distribution unstable and electron bunches are formed.

Phonon assistance designates two peculiar features of our system: first, it is
dissipative (not Hamiltonian); secondly, phonons brake phase relations for the
electron wave functions therefore nondiagonal elements of the density matrix
vanish and only probabilities of the site populations appear in consideration
and they give full and correct description of the system. If the energy
difference $\epsilon_{fi}<0$ (electron loses energy), there are two channels to
transit: first, it can undergo light-induced transition with the rate being
described in a way analogous to (\ref{w1}) with the only change $A$ to $-A$
(\cite{Kub}); second, it may relax to a lower energy state with the rate
$\gamma_{fi} =\gamma_{0}exp(-\kappa_{fi} R_{fi})$. So, for the electron
transition process with the energy loss we have:
\begin{equation} \label{w2}
w_{fi} = I \sigma_{0} cos^{2} \theta_{fj} exp(-\frac{(\hbar \omega -\epsilon_{if} + A)^{2} }{\Delta^{2} } -\kappa_{fi} R_{fi}) + \gamma_{0}exp(-\kappa_{fi}R_{fi}) \quad (\epsilon_{fi}<0)
\end{equation}
The factor $\gamma_{0}$ is of the order of the inverse lifetime for excited
electron states in an atom, so that $\gamma_{0} \approx 10^8 s^{-1}$. Light
driven kinetics is governed by dimensionless pumping constant $\mu =
\frac{I\sigma_0}{\gamma_0}$. Wave function of trapped electron outside the trap
$i$ is $\Psi_{i}({\bf R}) \propto e^{-\kappa_i |{\bf R-R}_i|}$, where
$\kappa_i= \sqrt{\frac{2m}{\hbar^2}(V-\epsilon_i)}$, $m$ is the electron mass,
$V$ is spacing of bottom of trapped levels from the bottom of the extended
states. Normal scale for this parameter is $\kappa_0 =
\sqrt{\frac{2m}{\hbar^2}V} \approx 10^{7}cm^{-1}$. Wave function of deep levels
vanishes rapidly outside the trap while shallow levels have long tails of wave
functions. Their overlap is determined by the slowest function, therefore
$\kappa_{fi}= \kappa_0 * min\{\sqrt{1-\frac{\epsilon_f}{V}},
\sqrt{1-\frac{\epsilon_i}{V}}\}$ (\cite{AntUFN}).

\begin{figure}
\includegraphics[width=16cm,height=14cm]{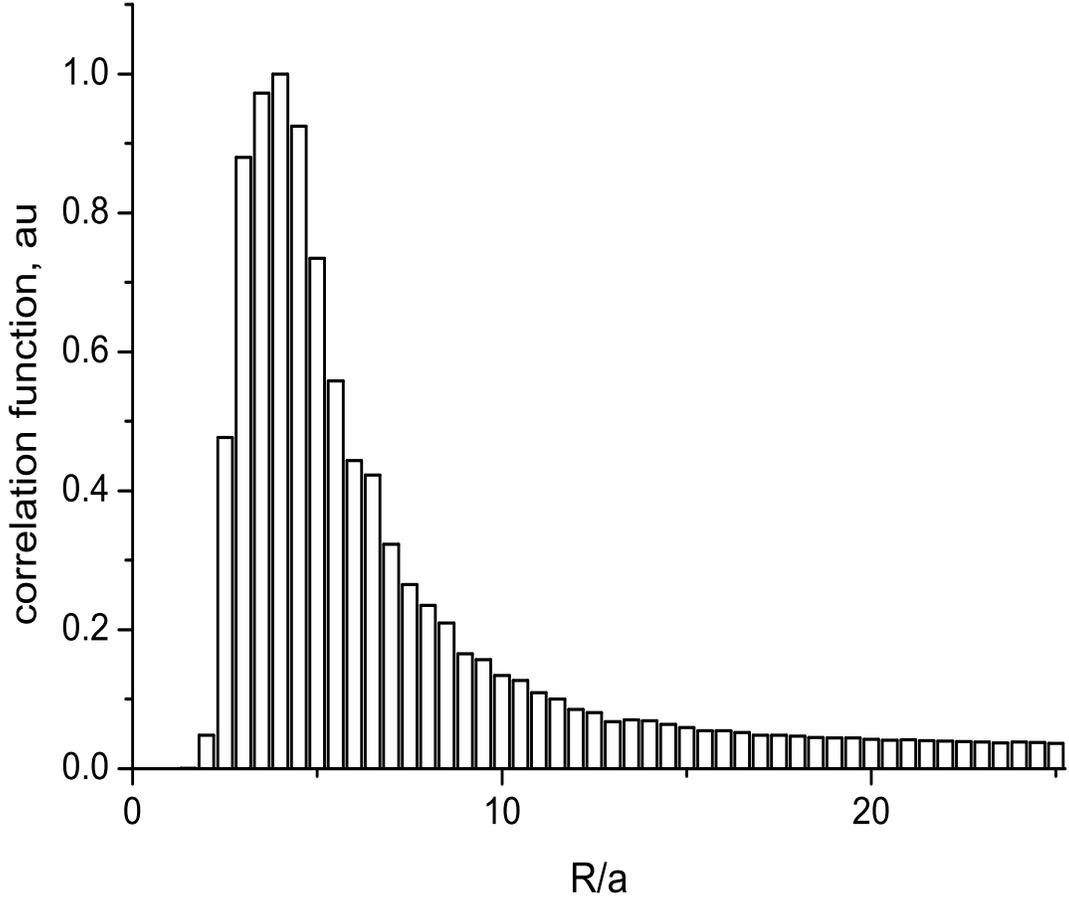}
\caption{Electron-electron correlation function revealing the bielectron state.
The parameters are: $\hbar \omega = 0.2 eV$, $\mu=100$, $V = 0.3 eV$, electron
density is constant within the range $0 \div V$.} \label{fig2}
\end{figure}

In the frame of the (\ref{w1}), (\ref{w2}) master rules we studied light driven
motion of two electrons and calculated their density-density correlation
function
\begin{equation}
K(R)= \frac{1}{TR^{d-1}}\int_0^T n({\bf R}_i)n({\bf R}_j)dt,
\end{equation}
where $R=|{\bf R}_i-{\bf R}_j|)$ and $n$ are the occupation numbers of the
corresponding traps, $t$ is time. We normalize the correlation function by the
factor $R^{d-1}$, where $d$ is dimensionality of the system. This normalization
avails to see the interaction effect: if the electrons were not interacting,
one would have $K(R)\equiv const$ for the noncorrelated case. The typical
$K(R)$ for three dimensions is shown in Fig. \ref{fig2}.

The bar chart displays the probability of different electron-electron ranges.
Electron traps were randomly distributed in space and single electron level
$\epsilon^{0}_i$ in each trap $i$ is also random according to some density of
states. One electron in position $j$ shifts each initial level $i$ of the
second electron to the energy of Coulomb interaction (and vice versa)
\begin{equation} \label{epsilon}
\epsilon_i = \epsilon^{0}_i + \frac{e^2}{R_{ij}}
\end{equation}
The parameters used for calculation are: average trap distance $a=5$ {\it \AA},
$\kappa_0 = 1/a$, $\hbar \omega = 0.2 eV$, impurity band width $V = 0.3 eV$,
$\Delta = 0.1 eV$, $A= 0.1 eV$. The bound state of two electrons is formed at
pumping constant $\mu=\frac{I\sigma_0}{\gamma_0} > 1$. Pumping constant $\mu =
1$ corresponds to laser power $10^7 W/cm^2$. We found that at high pumping in
wide region of parameters the correlation function has such a maximum. Despite
the Coulomb repulsion two electrons prefer to stay at some finite distance.
Analogous behavior has been observed for $d=2$ and $d=1$. Increase of average
distance $a$ between the traps (decrease of Coulomb interaction) results in
narrowing and vanishing of the peak in $K(R)$. At lower $\mu$ the peak in
$K(R)$ vanishes also (see for the details \cite{OptComm}).

Electron motion reveals competition between ordinary mobility in the direction
of the electric force $e\bf{E}$ given by relaxation term $\propto \gamma_{0}$
in (\ref{w2}) and light driven electron transfer in opposite direction. At long
distance between the electrons the light induced transitions dominate and an
electron is pushed to the other electron: light overcomes Coulomb repulsion
(!). This motion is stopped at short distances where Coulomb repulsion
dominates: electrons do not penetrate to short distance where correlation
function tends to zero (Fig. \ref{fig2}). Electric field at this distance is $E
= e/R^2 = 10^7 V/cm$ and this is an estimation for the effective electric field
caused by light induced electron transfer. This field is close to the damage
field $3*10^7 V/cm$. As the result of the competition bound state of two
electrons is formed with some preferable distance between particles. Effective
attraction between electrons results in light driven electron bunching. We
examined behavior of few electrons or even $10\div 100$ electrons and found
quite similar peculiarities. Energy shift of the initial levels for hopping
electron is analogous to (\ref{epsilon}) but is given by all other electrons:
\begin{equation}
\epsilon_i=\epsilon^0_i + \Sigma_j \frac{e^2}{R_{ij}}
\end{equation}

\begin{figure}
\includegraphics[width=16cm,height=14cm]{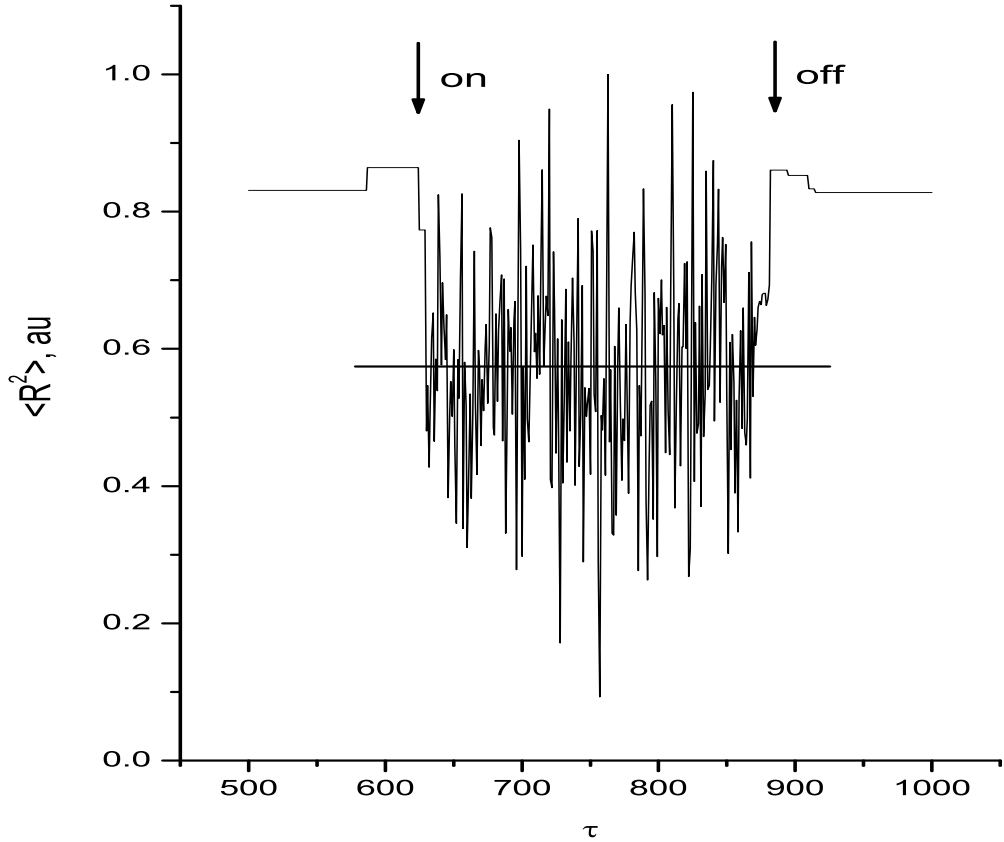}
\caption{Light driven bunching of the electrons. Squared size of the electron
cloud $<R^2>$ is decreased under light pumping (parameters are $E_0 = 0$,
$a=50$ {\it\AA}, $1/\kappa = 50$ {\it\AA}, $\mu= 100$, number of electrons $N =
10$.).}\label{fig7}
\end{figure}

\begin{figure}
\includegraphics[width=16cm,height=14cm]{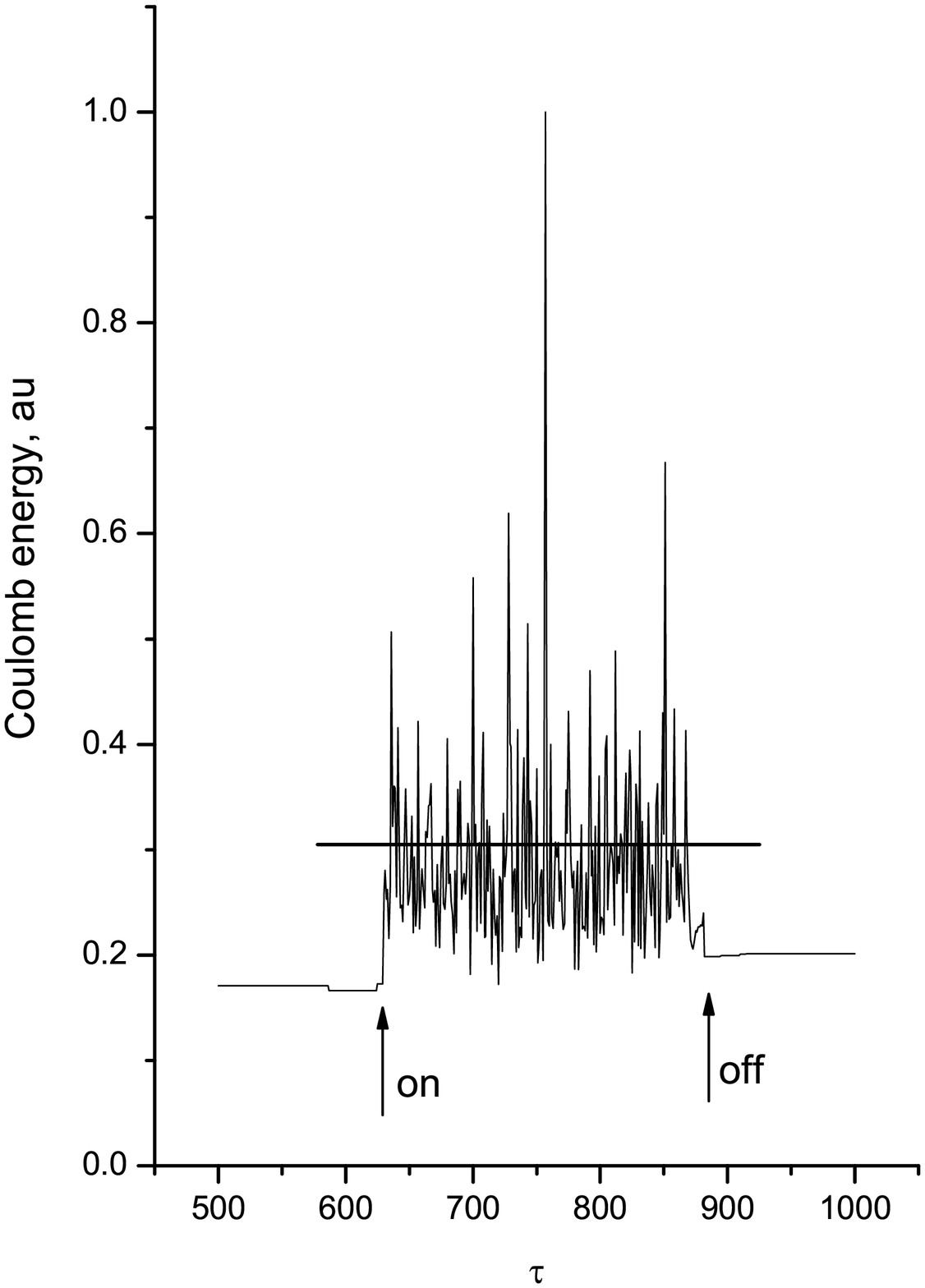}
\caption{Increase of electron Coulomb energy due to bunching under light
pumping (parameters are $E_0 = 0$, $a=50$ {\it\AA}, $1/\kappa = 50$ {\it\AA},
$\mu= 100$, $N = 10$).}\label{fig8}
\end{figure}

We studied temporal change of the average size of the electron cloud $R^2 =
1/2\Sigma_{ij}R^2_{ij}, R_{ij} \equiv |{\bf{R}}_i - {\bf{R}}_j|$ and Coulomb
energy $1/2\Sigma_{ij}e^2/R_{ij}$. The result is shown in Figs. \ref{fig7},
\ref{fig8}: electron cloud is pressed by light into compact bunch. We see again
electron motion which is impossible in thermodynamic case: the system tends to
the state with higher energy (Fig. \ref{fig8}). Bunch formation is accompanied
by strong increase of the electric field at the boundary of the bunch. The
formation is stopped by this field when it gains the value $\approx 10^7 V/cm$.
It is convenient to introduce ground state and electron-hole presentation for
the investigation of many body system in glass. Glass never attain full
thermodynamic equilibrium but partial equilibrium is possible and one may
consider that at ground state all electron levels bellow some energy
$\epsilon_F$ are occupied and states above this level are empty. Energy shift
of a level is determined in this case by the contribution of all electrons and
holes available.
\begin{equation}
\epsilon_i=\epsilon^0_i+\Sigma^e_j\frac{e^2}{R_{ij}}-\Sigma^h_j\frac{e^2}{R_{ij}}
\end{equation}

\begin{figure}
\includegraphics[width=14cm,height=12cm]{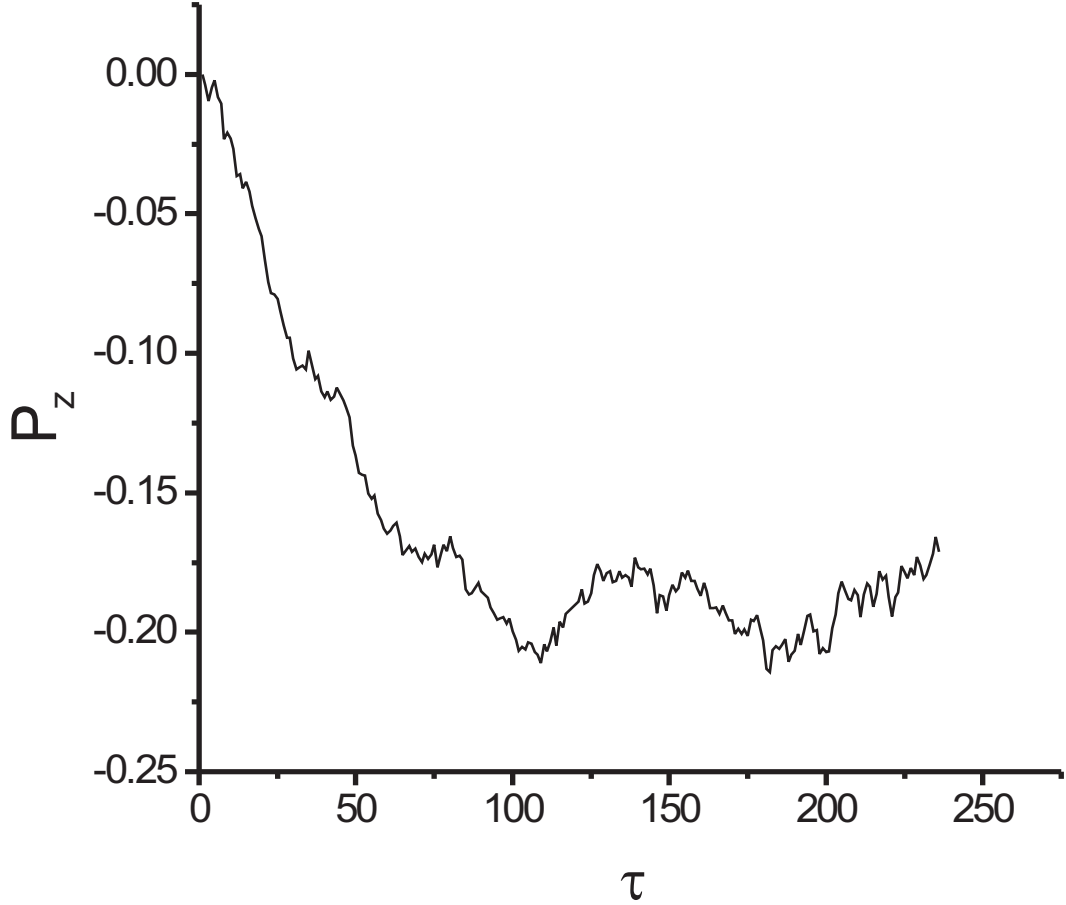}
\caption{Light induced polarization of the sample in $|e|/{a^{2}}$ units vs
dimensionless time $\tau = \gamma_{0} t$ for parameters $a=150$ {\it\AA}, $\mu=
0.001$, external field $E_0 \equiv Eea/\Delta=0.225$, $Ea^2/e = 0.02$, $\Delta
= 0.01 eV$.}\label{fig3}
\end{figure}

Contribution to the energy of the first excited electron is given by its "own"
hole only
\begin{equation}
\epsilon_i=\epsilon^0_i - \frac{e^2}{R_{ij}}
\end{equation}

where $j$ is position of the hole. Calculation of light induced polarization
have shown that it is directed in opposite to an external field direction (Fig.
\ref{fig3}) and therefore amplifies this field.

The studied three dimensional many body system with long range Coulomb
interaction reveals self-organization effect even at weak pumping $\mu \ll 1$.
We examined behavior of many body system of different size for one-, two- and
three-dimensional samples. In main features it is the same for all dimensions.
The corresponding model is presented in details in \cite{AntUFN},
\cite{OptComm}, \cite{Monograph}. Computer simulation of the electron kinetics
in the finite sample under external electric field shows that in agreement with
the above idea electrons are transported by light predominantly in the
direction opposite to the electric force (electric current flows contra the
voltage drop). This transfer is in contrast to the electron motion in a
thermodynamic case when electrons move in the direction of the external force
reducing in accordance with the Le Chatelet principle  the external field
(negative feedback). In our case glass reveals  positive feedback in response
to the electric field. As a result strong static polarization directed in
opposite to the external electric field is established. This polarization
amplifies the initial field considerably. At the field of saturation light
induced electron transitions are in the dynamical equilibrium with the ordinary
electron mobility in opposite direction. Computer simulation of the light
driven electron-hole kinetics have shown positive feedback in response to
static electric field in wide region of parameters independently on the
dimension of the system. This unusual response makes homogeneous electron
distribution unstable: an electron moves towards another electron or towards a
cluster of electrons and macroscopic bunches are formed.

\section{Experimental}

\begin{figure}
\includegraphics[width=14cm,height=12cm]{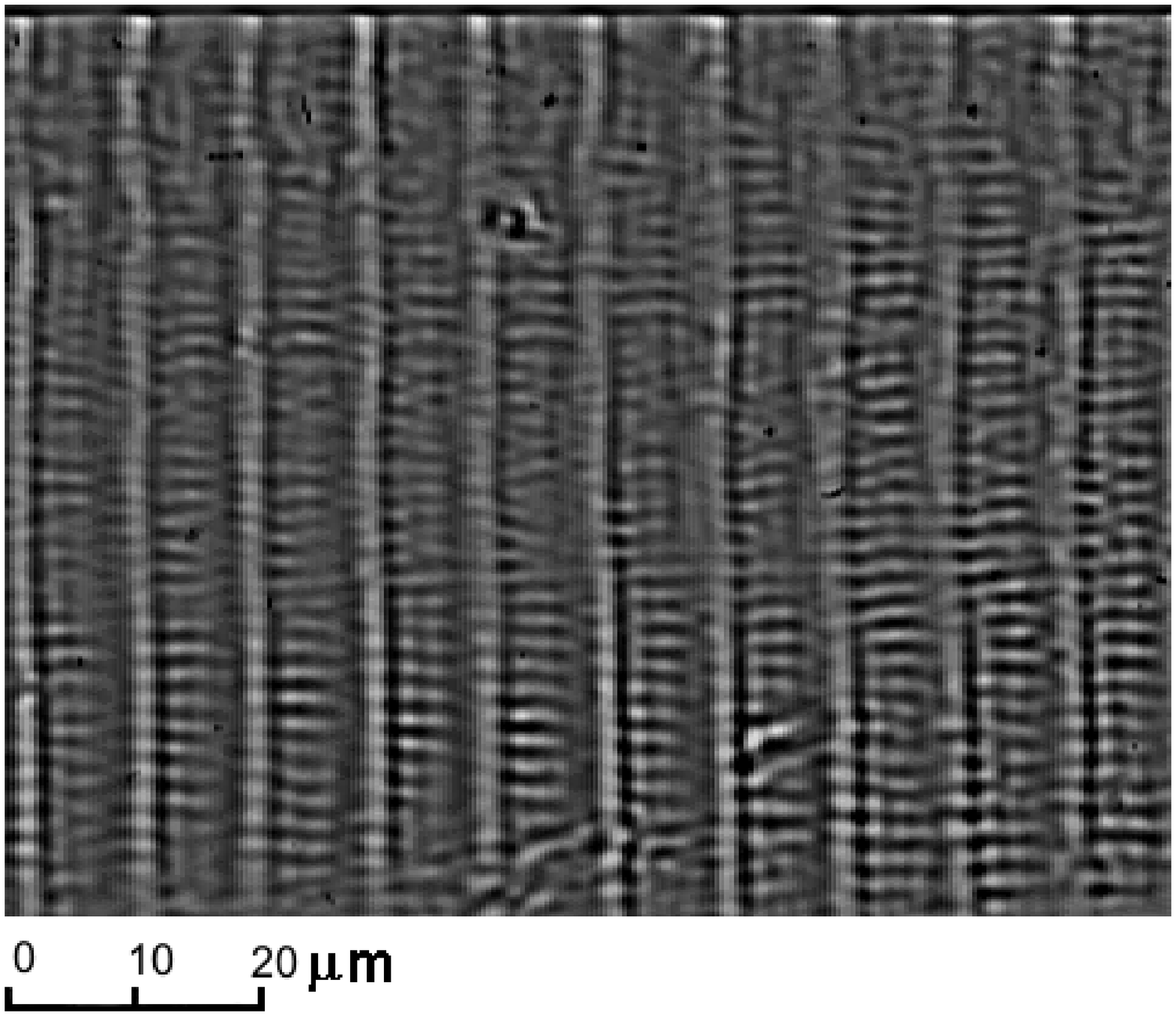}
\caption{Laser treated surfaces after ablation in two wave experiment: forced
grating is structured by self-organized system of domains with the period of
the bubble structure found in single wave experiment shown in Fig. \ref{fig12}.
Spatial coherence length exceeds 10 mm.}\label{fig13}
\end{figure}

We performed special experiment in order to observe directly the discussed
bunching. Strong laser wave with photon energy $\hbar \omega \approx 7 eV$ was
used. Band gap in our glass $\epsilon_g \approx 8 eV$ and therefore direct
generation of free electrons and holes is impossible. However in the region of
strong electric field due to Franz-Keldysh effect \cite{Franz}, \cite{Keldysh}
this process becomes effective. Energy discrepancy $\delta \epsilon$ between
the glass gap and photon energy is gained from the static electric field in
process of electron tunneling (the corresponding estimation see in
\cite{OptComm}). So, domain boundaries due to strong electric field become a
weak chain in glass bonds and light driven bond breaking (local fusion) reveals
bunch structure of the self-organized state. High quality UV-grade fused silica
samples (KU-1) with thicknesses of $1 \div 10 mm$ have been used in the
experiment. Before the experiment the samples have been subjected to ultrasonic
cleaning in acetone and ethanol followed by washing in de-ionized water. Two
beams from the unstable resonator of the $ArF$ excimer laser ($193 nm$,
CL-7000, PIC GPI) were focused on the rear side of a sample by the lens made of
$MgF_2$. We changed the angle between beams and therefore produced grating with
different periods at the output surface of the glass.  $ArF$ excimer laser
generated $20 ns$ FWHM pulses with energies up to $350 mJ$ and pulse repetition
rate up to $100 Hz$ ($\mu >> 1$). Ablation has been performed in ambient air as
well as in a hermetic chamber that was purged by nitrogen or evacuated. Silica
surface morphology and etched structures depth have been analyzed with an
optical microscope with further recording with Cohu-4812 CCD camera. Photograph
of the surface after the light treatment is presented in Fig. \ref{fig13}.

\begin{figure}
\includegraphics[width=16cm,height=12cm]{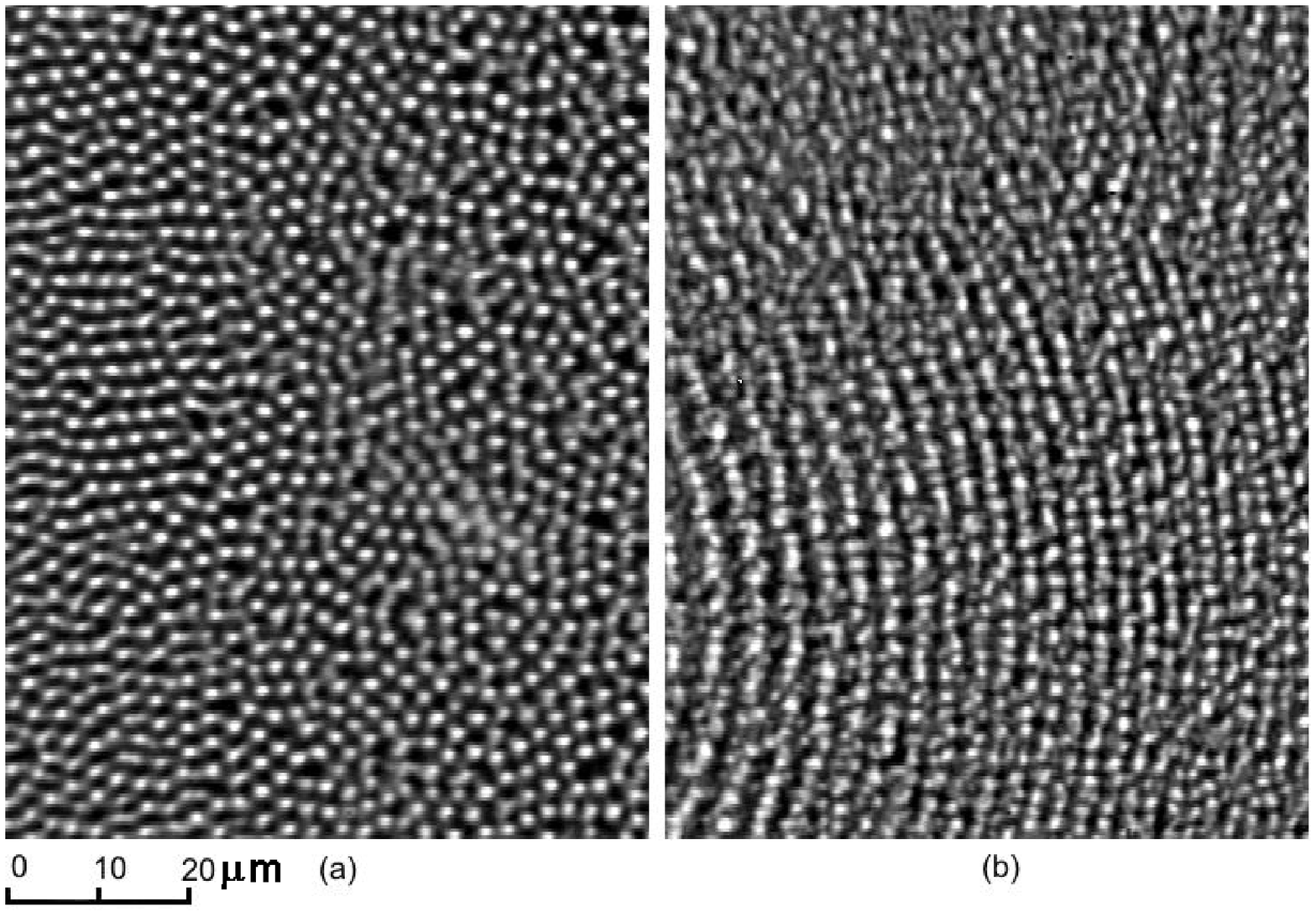}
\caption{Laser treated surfaces in single wave experiment \cite{OptComm}: a)
surface obtained after 300 pulses (laser fluence is $\approx 3 J/cm^2$) of
incoherent beam (spatial coherence length is less then $10 \mu m$; b) surface
obtained by highly coherent radiation treatment with spatial coherence length
$> 6 mm$ in all directions. Structures with the same spatial period $2 \mu m
\pm 0.2 \mu m$ are formed at different conditions a), b), and Fig.
\ref{fig13}.}\label{fig12}
\end{figure}

One can see farrows (grating) corresponding to the light intensity grating with
the period $\approx 10 \mu m$ and self-organized structure of each of them with
the period $2 \mu m \pm 0.2 \mu m$ corresponding to the discussed here
self-organized electron bunches. This size has nothing in common with the light
field scale (wavelength, beam size, coherent length etc.), it is determined by
internal properties of the glass what is characteristic feature of
self-organized system and coincides exactly with the period of the bubble
structure found in single wave experiment shown in Fig. \ref{fig12}
\cite{OptComm}. We changed coherent length of the light waves, beam sizes, beam
power and found that surface profiles (Figs. \ref{fig13}, \ref{fig12}) remain
the same: all experiments reveals the same spatial period $2 \mu m \pm 0.2 \mu
m$ what is typical for the self-organized systems. The material sputtering
(ablation) occurs on the rear side of a fused silica plate with no stimulation
of the process (such as plasma generation in the vicinity of a sample or
specific UV absorbing medium \cite{Jap}). Analogous behavior was observed in
\cite{interface} where surface structuring took place at the $SiO_2-Si$
interface and was explained in another way as interface property.

\section{Conclusions}

Our theoretical and experimental study of light driven electron kinetics in
silica glass have shown that homogeneous distribution of the trapped electrons
becomes unstable under light pumping and macroscopic electron (hole) bunches
are formed. In wide region of the external parameters light pushes electrons
into micron size coagulums which we observed in ablation process. Under
different conditions structures with the same spatial period $2 \mu m \pm 0.2
\mu m$ are formed. This new length is determined by the internal properties of
the material and is idependent on light properties (beam size, beam power,
coherent length etc.) It was found that UV beam $\lambda = 193 nm$ provides
bunching and subsequent local bond breaking effectively but wave $\lambda = 248
nm$ can not drive this process. Deeper electrons with short tails of electron
wave functions are involved in this case ($\kappa_{fi}$ is increased) and
probabilities of the transitions are less. In addition the energy discrepancy
$\delta \epsilon$ is increased therefore probability of Franz-Keldysh tunneling
becomes negligable. We observed also red luminescence of the glass sample under
UV pumping recently reported in number of papers (see \cite{Jap} and cited in
\cite{AntUFN} references). Ordering under light pumping discussed here may be
the basement of the so called "compaction effect" (see \cite{compaction} and
references therein) being the serious problem in lithography. Ordered by light
glass state certainly is more compact.

\section{Acknowledgements}

This work was performed under auspices of Russian Academy of Sciences in the
frame of the program "Optical spectroscopy and frequency standards" and Russian
Foundation for the Basic Research under Grant 04 -02-16394.

\end{document}